\documentstyle[11pt,twoside,jltp,amssymb]{article}

\title{Nonmonotonic Temperature Dependence of the Thermal Hall Angle of a
YBa$_{2}$Cu$_{3}$O$_{6.95}$ Single Crystal}

\author{R. Oca\~{n}a, A. Taldenkov$^{\dagger}$, P. Esquinazi, and Y. Kopelevich$^{*}$
\address{Department of Superconductivity and Magnetism, 
Institut  f\"ur Experimentelle Physik II, 
Universit\"at Leipzig, Linn\'estr. 5, D-04103 
Leipzig, Germany}}

\runninghead{R. Oca\~{n}a {\it et al.}}{Nonmonotonic Temperature 
Dependence of the Thermal Hall Angle}

\input psfig

\begin{document}

\begin{abstract}
We have performed high-resolution measurements of  the magnetic 
field (0~T~$ \le B \le 9~$T) and temperature (10~K~$\le T < 140~$K) dependence of the longitudinal 
and transverse Hall thermal conductivity of a twinned YBa$_{2}$Cu$_{3}$O$_{6.95}$ single crystal.
We have used and compared two recently published methods to extract the thermal Hall
angle $\theta_H(T,B)$. Our results indicate that $\cot(\theta_H)$
varies quite accurately as $T^4$ in the intermediate temperature range $\sim 0.3 < T/T_c$.
It shows a well defined minimum at $T_m \simeq 20~$K which resembles that observed 
in the $c$-axis microwave conductivity. The electronic part of the longitudinal and 
the transverse thermal conductivity show the scaling
behavior for transport properties predicted for d-wave superconductors in the
 temperature range $\sim 18~$K~$\le T \le 30~$K.

PACS numbers: 74.25.Fy,74.72.Bk,72.15.He

\end{abstract}

\maketitle
\vspace{0.3in}

\section{Introduction}
A characteristic feature of the high-$T_c$ superconductors (HTS) is the $T^{-2}$ temperature
dependence of the normal state Hall angle $\tan(\theta_H)$ which has been explained by
Anderson  within the framework of the Luttinger liquid theory \cite{anhall}. Recently
performed thermal Hall conductivity measurements on YBa$_{2}$Cu$_{3}$O$_{x}$ (Y123) single crystals
at $T < T_c$ suggest  that such a behavior would extend down to $\sim 50~$K \cite{zeini,kris2}.
Obviously the $T^{-2}$ dependence cannot continue to arbitrarily low temperatures since
it would imply an infinite scattering time or mean free path for the quasiparticles.
Nevertheless, it is unclear from literature which is the temperature dependence of the Hall 
angle below $T_c$ and at which temperature the monotonic temperature
dependence breaks down. Apart from a possible saturation of $\tan(\theta_H) \propto l^{\ast}(T)$ 
(an effective mean free path of quasiparticles) at low enough temperatures, 
there are at least two more reasons to expect a deviation 
from a monotonous temperature dependence at low temperatures. (1)
Theory predicts a maximum in $\theta_H(T)$ at low enough temperatures due to a
crossover from holon non-drag regime to  localization \cite{anhall2}. (2) Electrical transport 
measurements performed on underdoped YBa$_2$Cu$_3$O$_{6.63}$ \cite{xu} and
 Zn-doped YBa$_2$Cu$_3$O$_{7-\delta}$ \cite{abe} crystals as well as 
in Bi$_2$Sr$_2$Ca$_{n-1}$Cu$_n$O$_y$ 
thin films \cite{konst}suggest  
that the temperature dependence of the (electrical)
Hall angle $\tan(\theta_H) \propto \tau_H/m_H$ ($\tau_H$ is the Hall scattering time and $m_H$ the
effective mass of the quasiparticles responsible for the Hall signal) has a maximum 
at or near the temperature at which the pseudogap opens. 
Experimental evidence accumulated over the last years suggests also that the opening of a pseudogap 
can take place below the superconducting transition temperature affecting the electronic properties of 
HTS \cite{talo,bat}, although its nature and its relationship to superconductivity are not yet well
understood. Whereas the electrical Hall angle can hardly be measured
at $T \ll T_c$, the thermal transport is suitable for the low temperature studies.

Below the superconducting critical temperature $T_c$ and decreasing temperature, 
the in-plane longitudinal thermal conductivity $\kappa_{xx}$ in HTS increases and shows
a maximum between  $0.3 T_c < T < 0.9 T_c$.  This behavior is
attributed nowadays mainly to the increase 
of the electronic contribution $\kappa_{xx}^{\rm el}(T)$ due to the
increase in the quasiparticle-quasiparticle relaxation time $\tau$.
It appears that the increase in $\tau$ decreasing
$T$ overwhelms the decrease of the density of quasiparticles due to their condensation
in the  superconducting state. Pioneer work on the thermal conductivity of  YBa$_2$Cu$_3$O$_7$ 
(Y123) crystals \cite{yu1} as well as thermal Hall effect measurements \cite{krishana,harris,zeini,kris2}
provide strong evidence for this interpretation. 
Regarding the difference between the Hall $\tau_H(T,B)$ and diagonal $\tau(T,B)$ relaxation  times we note
that whereas $\tau$ shows a monotonic enhancement decreasing $T$ below $T_c$, $\tau_H/m_H$ appears to be
unaffected by the superconducting transition \cite{zeini,kris2}. 

The difficulty to describe the behavior of the thermal conductivity and to obtain from the
experimental data the temperature and field dependence of the Hall angle
resides mainly in the method to separate the electronic contribution 
$\kappa_{xx}^{\rm el}$ from the measured total thermal conductivity.
This is due to the relatively large phonon contribution to the thermal transport in HTS.
Basically two methods for this separation have been treated in the literature. One method is
based on a phenomenological description of the field dependence of $\kappa_{xx}(T,B)$, 
introduced  first by
Vinen et al. \cite{vinen} and used in Refs.~ \cite{pogo,yu-kopy,kris2} to estimate $\kappa_{xx}^{\rm el}(T)$. 
The other, apparently more elegant method presented by Zeini et al. \cite{zeini}, 
is based on simultaneous measurements of the longitudinal and transverse thermal
conductivity and the assumption of a field independent Hall relaxation time. 

We note that the experimental error of  the published 
data of the thermal
Hall angle leaves its true temperature and field dependence not well defined 
below $T_c$.  In this work
we obtain the temperature dependence of  $\theta_H(T,B)$ down to $\sim 10$~K using high-resolution
experimental data of  $\kappa_{xx}$ and $\kappa_{xy}$ and the two recently proposed separation methods.
We will show that the ratio $m_H / \tau_H$ follows a $(T/T_c)^4$ dependence that agrees with
the temperature dependence of the scattering rate obtained from longitudinal thermal conductivity
measurements.

\section{Sample and experimental details}
For the measurements we have used a twinned, nearly optimally doped
YBa$_{2}$Cu$_{3}$O$_{6.95}$ single crystal
with dimensions (length $\times$
width $\times$ thickness)~$0.83 \times 0.6 \times 0.045~$mm$^3$ and $T_c = 93.4~$K. 
The use of a twinned crystal, with twinning planes parallel and perpendicular to the
heat current, allows us to apply both separation methods described below, as well as
to rule out the influence of orthorhombicity on the heat transport properties.
The temperature
and field dependence of $\kappa_{xx}$ for this crystal have been recently measured below and above $T_c$ 
with a relative accuracy of $10^{-4}$ \cite{tal}. The transverse thermal conductivity has been determined
using the relation $\kappa_{xy} = \kappa_{xx} \Delta_y T/ \Delta_x T$, where $\Delta_{x} T$  is the applied 
thermal gradient in $x$ direction and $\Delta_y T$ is perpendicular to it when a magnetic field
is applied in the $z$ or $-z$ direction (parallel to the $c$-axis of the crystal). Because of twinning we
have assumed that $\kappa_{xx} = \kappa_{yy}$. The temperature
gradients were measured with a previously field- and temperature-calibrated
type E thermocouples \cite{iny}. Their voltages were measured with a dc picovoltmeter 
which allowed a resolution of $\sim 30~\mu$K at 100~K ($\sim 60~\mu$K at 10~K). The short time 
(2 h) temperature stability of the sample holder was $50~\mu$K,
for more details see Refs.~\cite{tal,iny}. 

In all the temperature range (10~K~$\le T \le 140~$K) we have applied relatively small temperature
gradients along the sample, typically $\Delta_x T \le 300~$mK, in order to
diminish smearing effects in the $T$-dependence of the measured properties. 
$\Delta_y(T,B)$ was determined from the difference $[\Delta_y(T,B) - \Delta_y(T,-B)]/2$ 
to eliminate offset contributions.
The thermal Hall effect was measured as a function of temperature at constant $B$ and $-B$
in the field-cooled state of the sample in order to rule out pinning effects \cite{tal}. 
The influence of the pinning of vortices to the Hall effect in the mixed state of
superconductors should be  considered seriously and carefully minimized, specially
at low temperatures \cite{gil}. We have checked 
the results measuring also
the field dependence of $\kappa_{xy}$ at constant selected temperatures. 
We have obtained very good agreement
between both methods.

\section{Results}

\begin{figure}
\centerline{\psfig{file=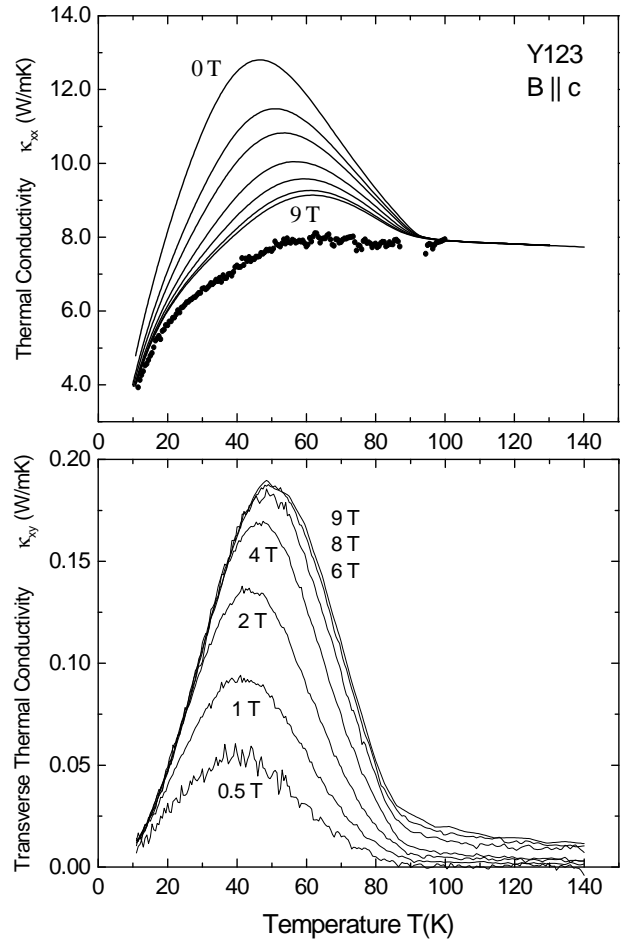,height=5.5in}}
\caption{Longitudinal (upper figure) and transverse (lower figure) thermal conductivity
as a function of temperature at constant applied fields. The curves in the upper figure (from top to
bottom) were obtained at $B = 0, 1, 2, 4, 6, 8, 9$~T. The close symbols represent the
phonon or rest contribution to the total thermal conductivity.}
\label{dataset}
\end{figure}

\begin{figure}
\centerline{\psfig{file=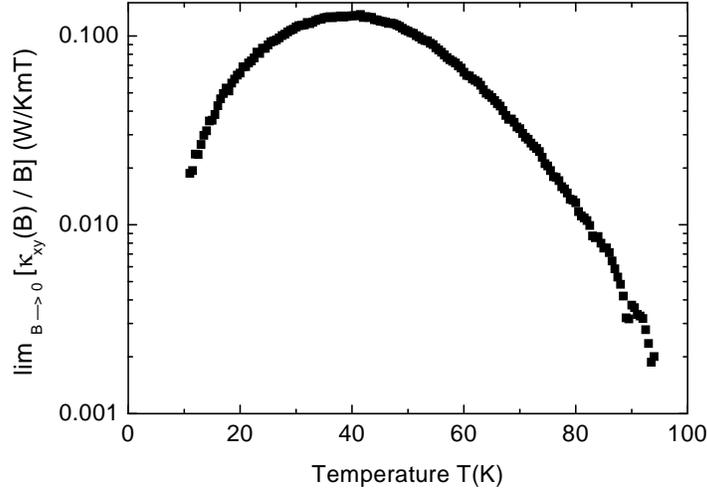,height=3.0in}}
\caption{Initial Hall slope $\lim_{B\to 0} \kappa_{xy}/B$ as a function of temperature.}
\label{lim}
\end{figure}

Figure \ref{dataset} shows the longitudinal $\kappa_{xx}$ and transverse
$\kappa_{xy}$ thermal conductivity as a function of temperature at
different constant applied fields. The overall $T-$dependence as well as
the absolute value of those properties agree well with published results
\cite{harris,zeini}. From these data  we
may obtain $\tau_H/m_H$ if the electronic contribution $\kappa_{xx}^{\rm el}(T)$ is known since according to
the usual definition
\begin{equation}
\frac{\tan(\theta_H)}{B} = \frac{e \tau_H}{m_H} = \frac{\kappa_{xy}}{B \kappa^{\rm el}_{xx}}\,.
\label{tan}
\end{equation}
As in Ref.~\cite{zeini} our data also show that, in good approximation, 
$\partial (\kappa_{xy}/B)/\partial B \propto \partial \kappa_{xx}/\partial B$
in the whole measured temperature range. This fact was used by Zeini et al. \cite{zeini}
to estimate the Hall relaxation time assuming that $\kappa_{xy}$ and $\kappa_{xx}^{\rm el}$
have a similar field dependence, neglecting the field dependence in $\tau_H/m_H$ or in the rest
contribution to the thermal conductivity. Following 
Ref.~\cite{zeini} for a pair of values of field $(B_i,B_j)$ with $B_i \neq B_j$ we can
write
\begin{equation}
\frac{\kappa_{xy}(B_i)}{B_i} -\frac{\kappa_{xy}(B_j)}{B_j} = 
\frac{\tan(\theta_H)}{B} (\kappa_{xx}(B_i) - \kappa_{xx}(B_j))\,.
\label{tanzeini}
\end{equation}
Taking pairs of points at low fields we can also calculate the initial Hall slope $\lim _{B \to 0} \kappa_{xy}/B$.
This quantity is depicted in Fig.~\ref{lim}. We note that $\lim _{B \to 0} \kappa_{xy}/B$ increases
two orders of magnitude below $T_c$ reaching a maximum at 40~K.
\begin{figure}
\centerline{\psfig{file=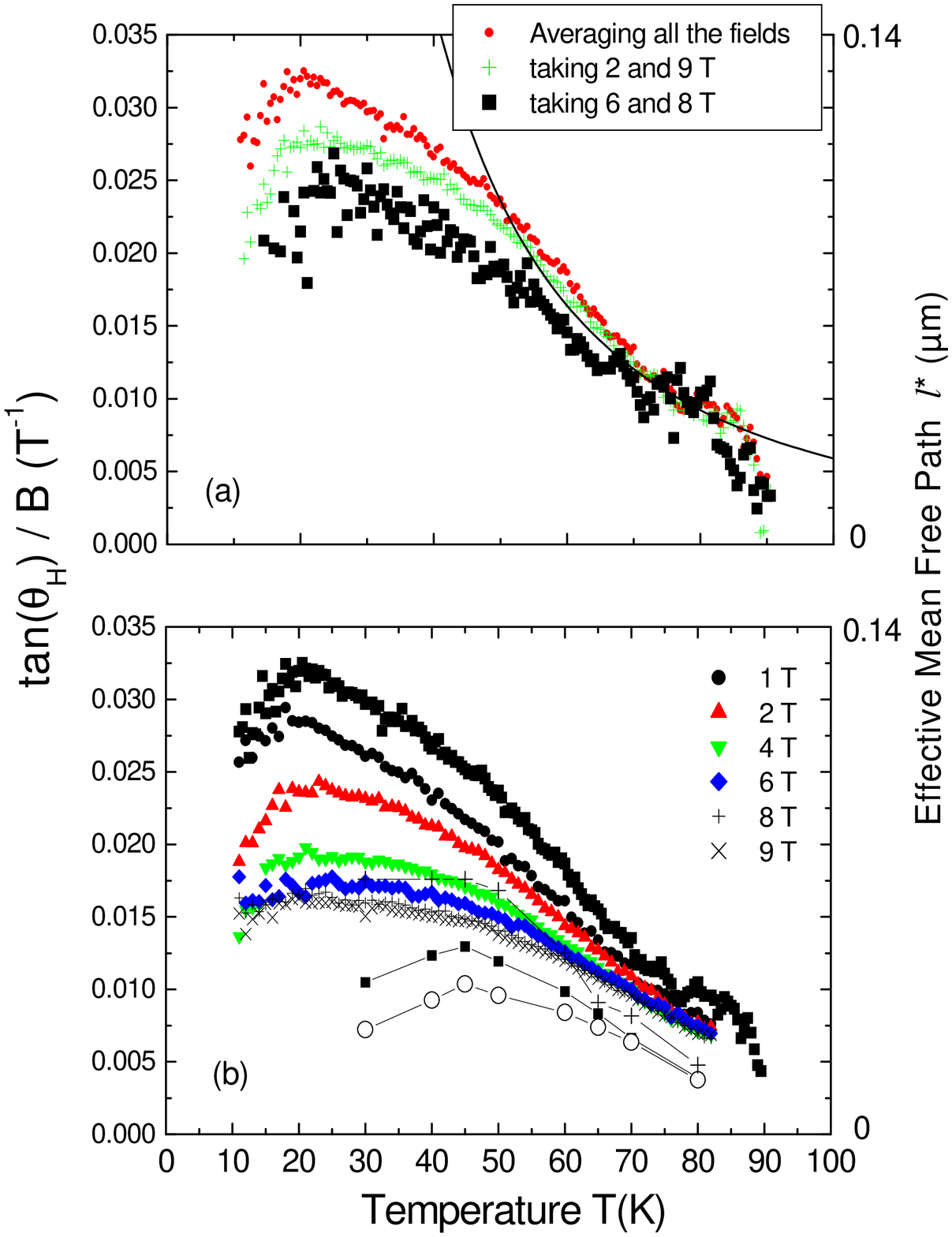,height=5.2in}}
\caption{(a) The ratio $\tan(\theta_H)/B = e \tau_H/m_H$ ($\theta_H$ is the Hall angle, $e$ the electron charge, and
$m_H$ the mass of the particles responsible for the Hall effect) as a function of temperature obtained assuming
a field independent $\tau_H$ within the approach of Ref.~2, see Eq.~(\protect\ref{tanzeini}). 
The three curves
are obtained averaging the pairs at all measured magnetic fields ($\bullet$), taking only the values
at 2 T and 9 T ($+$), and at 6 T and 8 T ($\blacksquare)$. The continuous line has a $T^{-2}$ dependence.  
(b) The same as (a) at different applied fields calculated
 using $\kappa_{xx}^{el}$ obtained by fitting 
the field dependence of the longitudinal thermal conductivity. 
$(\blacksquare)$: Data from (a) averaging all the fields.
The three curves
with points connected by straight lines are taken from Ref.~3 at $B = 2~$T $(+)$, 6~T~$(\blacksquare)$ 
and
10~T~$(\bigcirc)$. The right axis shows the scale of the calculated effective mean free path in $\mu$m.}
\label{figtan_s}
\end{figure}

In Fig.~\ref{figtan_s}(a) we show $\tan(\theta_H)/B$ obtained using (\ref{tanzeini}) 
for two pairs of fields and also from a linear regression considering  the data at all fields.
The results in Fig.~\ref{figtan_s}(a) indicate that $\tau_H/m_H$ clearly deviates from a $T^{-2}$
dependence and shows a maximum at $T_m \simeq 20~$K. 

The results in Fig.~\ref{figtan_s}(a) also indicate that
$\tan(\theta_H)/B$ depends slightly on the pairs of fields used to compute it and 
decreases the larger the field  of the chosen pair. It appears that the values obtained
from the linear regression would provide $\tan(\theta_H)/B$ at $B \rightarrow 0$.
The Zeini et al. approach is based on the
assumption of a field independent $\tau_H/m_H$ or, in other words, a strictly linear field
dependence for $\tan(\theta_H)$. Deviation from this assumption can be proved using the second separation
method \cite{pogo,kris2} to obtain $\kappa_{xx}^{\rm el}$ from the experimental
data. Therefore, we use the second separation method in order to show that the
temperature dependence of $\tan(\theta_H)$ at low enough fields is independent of the assumption of a
field independent $\tau_H/m_H$.

As it was shown recently \cite{tal}, in agreement with Ref.~3, the field dependence 
of  $\kappa_{xx}$ (for $B || c$-axis) at all temperatures below $T_c$ can be very well fitted assuming
\begin{equation}
\kappa_{xx}(T,B) = \kappa_{xx}^{\rm ph}(T) + \frac{\kappa_{xx}^{\rm el}(T)}{1 + \beta_e(T) B}\,,
\label{kapa}
\end{equation}
where $\beta_e(T) $ is proportional to the zero-field electronic mean-free-path (or the longitudinal
relaxation time) of the quasiparticles \cite{krishana,pogo}.
Fitting the field dependence of $\kappa_{xx}(B)$ at different constant temperatures we can separate
the electronic contribution and calculate $\tan(\theta_H)/B$ as a function of temperature at different
applied fields. These results are shown in Fig.~\ref{figtan_s}(b). At low fields, the temperature dependence of
$\tan(\theta_H)/B$ resembles that obtained with the other separation method,  specially
 the maximum at $\sim 20~$K, compare Figs.~\ref{figtan_s}(a) and (b). 
We see also clearly that $\tan(\theta_H)/B$
decreases with  field in the whole temperature range below $T_c$. 
Note that at high enough fields the maximum at $T \simeq 20~$K vanishes.
Our data agree reasonably well with those we get using
the data from Ref.~\cite{kris2}, see Fig.~\ref{figtan_s}(b).  We note that the data of  Ref.~\cite{kris2} 
show a maximum in $\tan(\theta_H)/B$ at $T \sim 45~$K and at high fields.

It appears that the results obtained 
using the Zeini et al.
approach \cite{zeini}
are similar to those obtained  with the Krishana et al.\cite{kris2} for $B \rightarrow 0$. 
The field dependence
of $\kappa_{xy}(B)$ is given in good approximation  by the function $B / (1 + \beta_H B)^2$ 
where $\beta_H(T)$ is
a temperature dependent constant analogous to $\beta_e(T)$, see Eq.~(\ref{kapa}). 
Because
$\tau_H/m_H \propto \kappa_{xy}/(B \kappa_{xx}^{el})$ and, in principle, $\beta_e(T) \neq \beta_H(T)$,
 the field dependence of $\tau_H/m_H$ can be neglected in the limit $B \rightarrow 0$. 
It seems therefore reasonable that in good approximation  both approaches provide similar
results for the ratio $\tan(\theta_H)/B$
at $B \rightarrow 0$. 

A rough relationship between $\tan(\theta_H)/B$ and the effective mean free path
of quasiparticles can be obtained if we assume that $\tan(\theta_H) \sim \omega_c \tau =
e B \tau/m^\ast \sim eBl/m^\ast v_F \sim e B l^\ast /\hbar k_F$. Here $v_F$ and $k_F$ are the
Fermi velocity and wave vector, and $m^\ast$ the effective mass of quasiparticles. Using
$k_F \sim 0.6 \times 10^{10}~$m$^{-1}$ we calculate the effective mean free path shown
in the right axis scale of Fig.~\ref{figtan_s}.

\begin{figure}
\centerline{\psfig{file=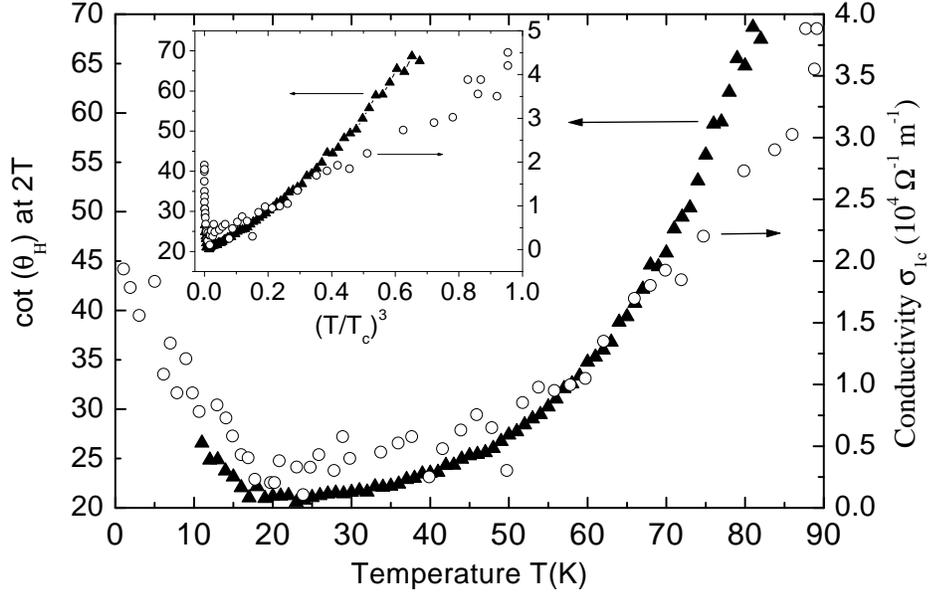,height=3.7in}}
\caption{Temperature dependence of  $\cot(\theta_H) \propto m_H/\tau_H$ obtained at 
$B = 2~$T ($\blacktriangle)$ and the $c$-axis
microwave conductivity ($\bigcirc$, right axis) taken from Ref.~\protect\cite{hos}. 
Note that the origins of the vertical axes differ.
The inset shows the same data but as a function of $(T/T_c)^3$.}
\label{fig_tau}
\end{figure} 

Figure \ref{fig_tau} shows the temperature dependence of $\cot(\theta_H)$
obtained from the data in Fig.~\ref{figtan_s}(b) at $B = 2~$T. 
We stress that $\cot(\theta_H)$ does not follow the $T^2$ dependence observed at 
higher temperatures \cite{zeini,kris2} and 
shows a minimum at $T_m \simeq 20~$K. We note further that the measured temperature dependence of 
$\cot(\theta_H)$ shows a striking similarity
with that measured for the $c$-axis microwave conductivity obtained at 22 GHz in 
a Y123 crystal with similar $T_c$ \cite{hos}, 
see Fig.~\ref{fig_tau}. From Ref.~\cite{kris2} we found $T_m \sim 45~$K (see Fig.~2) suggesting
that $T_m$ varies from crystal to crystal. 
A comparison between our results 
\begin{figure}
\centerline{\psfig{file=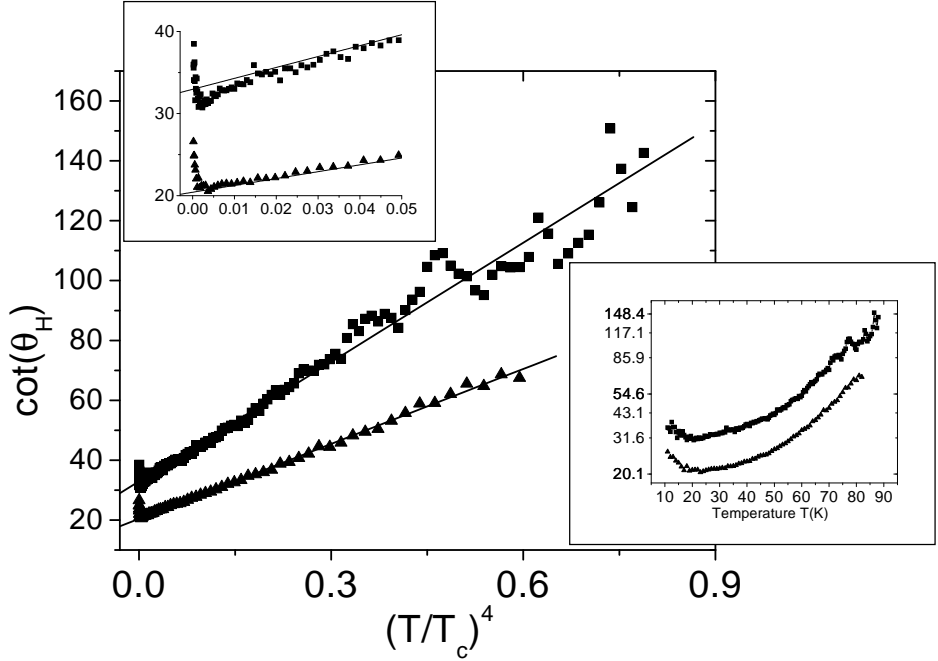,height=3.5in}}
\caption{$\cot(\theta_H) \propto m_H/\tau_H$ obtained at 
$B = 2~$T ($\blacktriangle)$ (from Fig.~\protect\ref{figtan_s}(b)) and
averaging all the fields $(\blacksquare)$ (from Fig.~\protect\ref{figtan_s}(a) 
and assuming $B = 1~$T) as 
a function of $(T/T_c)^4$. The upper left inset shows the same data at the lowest measured
 temperatures. The straight lines are linear fits taking into account all the points for
$T > T_m$. The botton right inset shows the same data but in a semilogarithmic scale.}
\label{t4}
\end{figure} 
and those from Ref.~\cite{kris2} indicates that at high enough fields
the difference in the density of scattering centers between samples does not influence
the absolute value of $\theta_H/B$  significantly, see Fig.~\ref{figtan_s}.
It is clear that to demonstrate the possible influence of the sample purity on the mean free path
the thermal Hall angle should be obtained at the low-field limit.
We note also that  the temperature of the minimum in $\sigma_{1c}(T)$ can be shifted
from $\sim$20~K to $\sim$40~K depending on the crystal \cite{hos}. 
We will discuss in the next section possible origins of the upturn in $\cot(\theta_H)$.

In a recently published paper \cite{xiang} 
it was found that the $c$-axis conductivity of the crystal reported in 
Ref.~\cite{hos} follows approximately 
a $T^3$ dependence below $T_c$ and for $T > T_m$, see inset in Fig.~\ref{fig_tau}.
This behavior is shown to be consistent with an anisotropic 
interlayer hopping integral.\cite{xiang} The authors in Ref.~\cite{xiang}
further show that the observed temperature dependence of the $c-$axis conductivity in the
intermediate temperature range should be an universal result independent of
the scattering rate of quasiparticles. Since $\cot(\theta_H)$ is directly proportional
to the scattering rate of the quasiparticles responsible for the Hall signal and assuming that the
result of Ref.~\cite{xiang} is valid, it would be rather
surprising if both quantities,  $\cot(\theta_H)$ and $\sigma_c$,  show the same 
temperature dependence. In fact,
we show in the inset of Fig.~\ref{fig_tau} that $\cot(\theta_H)$ does not follow
a $T^3$ dependence. Our results show that $\cot(\theta_H)$ follows
quite accurately a $T^4$ dependence above $T_m$ and that this dependence is independent
of the separation method used, see Fig.~\ref{t4}. The inset at the botton of Fig.~\ref{t4} shows also  
that $\cot(\theta_H)$ does not follow an exponential law, $\exp(T/T_0)$, as obtained for the
quasiparticle scattering rate from
microwave surface resistance using different assumptions.\cite{bonn} 
To the best of our knowledge this  is the first time that a clear power-law 
dependence $(T/T_c)^4$ in a broad temperature range is
reported for the Hall scattering rate below $T_c$.
This result and the minimum at $T \simeq 20~$K are the main messages of the 
present work. 

The temperature dependence of the quasiparticle scattering rate below $T_c$ is
not known at present. From longitudinal thermal conductivity
measurements and assuming a $d-$wave pairing Yu et al. \cite{yu1} obtained a scattering rate 
that follows roughly a $(T/T_c)^4$ dependence in agreement with our result.
On the other hand, since 
quasiparticle-quasiparticle scattering is the dominant temperature dependent 
scattering mechanism in
our sample, we expect a scattering rate proportional to the density
of quasiparticles. In the simple two-fluid model and for an isotropic order parameter
we expect a density of quasiparticles proportional to $(T/T_c)^4$.\cite{tink}

\section{Discussion}

In this section we restrict ourselves to discuss the nonmonotonic 
behavior obtained for the Hall angle and its  scaling properties.
 Note first that a simple background scattering contribution to the effective
quasiparticle mean free path should saturate $l^\ast(T)$ at low enough
temperatures. However, that $\tan(\theta_H)/B$ shows a nonmonotonic 
temperature dependence is a non-trivial result not yet clearly reported
in the literature. The reason for this temperature dependence is not known 
at present. Whatever its origin, the results indicate that some
additional scattering, change in the quasiparticle density and/or a change
in the effective mass 
takes place below $\sim 20~$K in our sample. It would be interesting
to see whether our results follow the scaling relations
for  transport properties of d-wave superconductors proposed in
the last few years and  whether this additional scattering/quasiparticle 
density or effective mass change has some effect or not. 

Volovik and Kopnin \cite{vol1,kop} showed that in unconventional superconductors 
and because of the presence of low energy excitations associated with the gap nodes, 
at low temperatures and low fields $(T \ll T_c, B \ll B_{c2})$ the thermodynamic and 
transport properties are dominated by the influence of the Doppler shift on the excitation 
spectrum of the quasiparticles due to the supercurrents flowing around the vortex cores. 
Simon and Lee \cite{sim1} predicted a scaling function with only one dimensionless 
parameter at temperatures below $\sim 30~$K such that for the heat capacity 
$C(T,B)/T^2 \propto f(x)$ and the Hall component of the thermal conductivity $\kappa_{xy}/T^2
\propto f_{xy}(x)$ with $x=\alpha \sqrt{B}/T$. The function $f$ and $f_{xy}$ are universal
functions. The coefficient $\alpha$ should be equal to $T_c/\sqrt{B_{c2}}$ according to 
Refs. \cite{vol2,sim2} to preserve the correct  predicted regimes. Scaling 
relations can only be extrapolated to the Hall (off-diagonal) component of the thermal 
conductivity tensor because only this component is electronic in origin. In principle, 
this picture may also be valid for systems with a 3D order parameter with line nodes such as in UPt$_3$ 
as has been shown experimentally \cite{sud}. 
For Y123 crystals the scaling model predicts two changes of regime 
at the following points: (1) at 
$x_1 \sim E_F/T_c$ (i.e. at $\sqrt{B}/T \sim  2.18$~ 
Tesla$^{1/2}$/K) where the discreteness of the fermion bound states in the vortex 
becomes important (the quasiclassical approach does not hold longer) and 
(2) at $x_2 \sim1$ (i.e. at $\sqrt{B}/T \sim 0.068~$Tesla$^{1/2}$/K) where the single vortex contribution 
is comparable to the bulk contribution per one vortex \cite{vol2,sim2}.

\begin{figure}
\centerline{\psfig{file=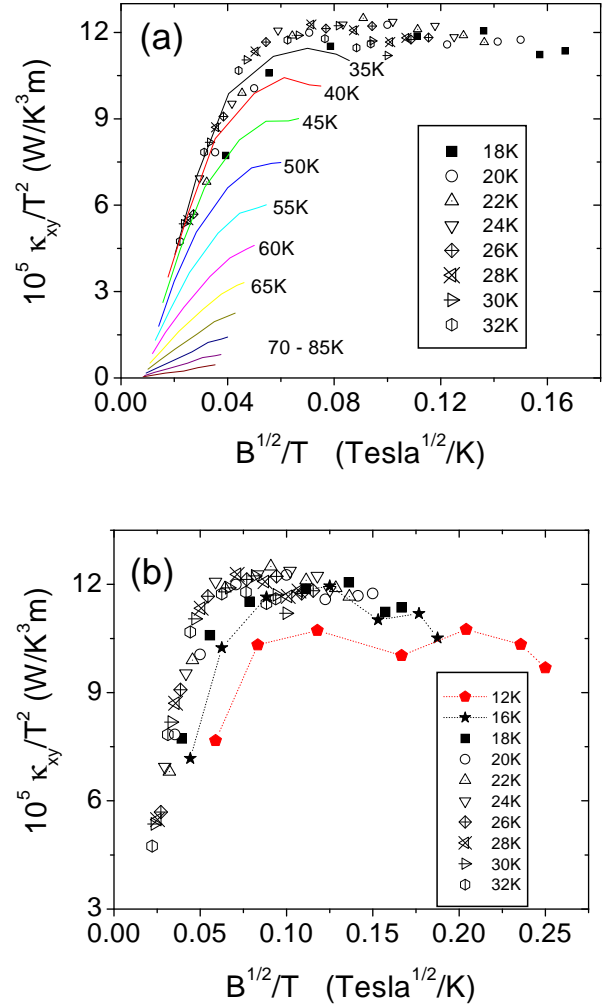,height=6.0in}}
\caption{Hall thermal conductivity divided by the square of the temperature as
a function of the scaling variable $B^{1/2}/T$ at different fixed temperatures:
(a): 18~K~$\le T \le~$85~K; (b):  12~K~$\le T \le~$32~K.}
\label{sca}
\end{figure}

\begin{figure}
\centerline{\psfig{file=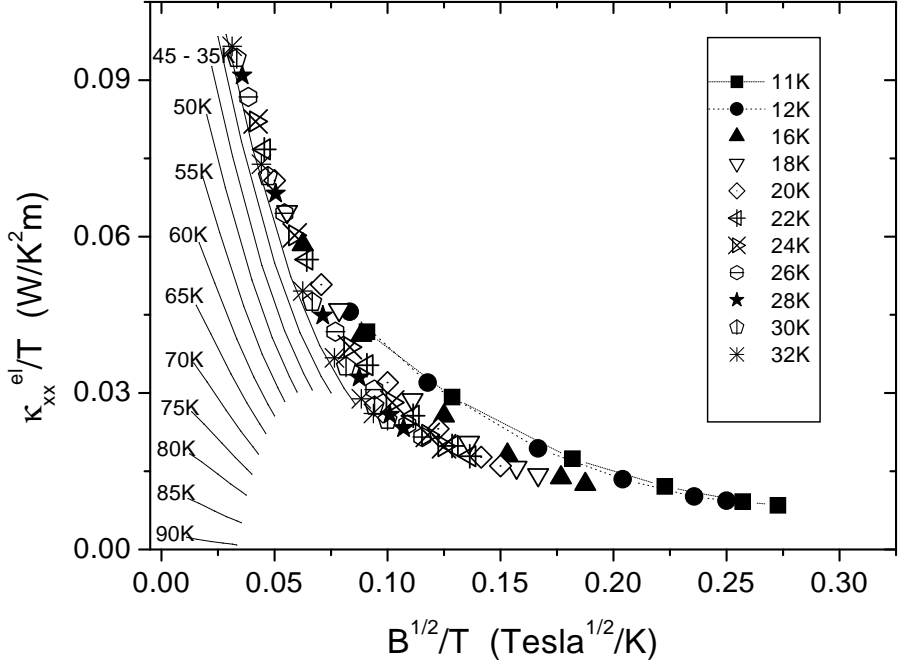,height=4.0in}}
\caption{Longitudinal electronic thermal conductivity divided by the temperature as
a function of the scaling variable $B^{1/2}/T$ at different fixed temperatures.
At 18~K~$\le T \le~$32~K $\kappa^{el}_{xx}$ shows the predicted scaling behavior.}
\label{keescal}
\end{figure}

  In Fig.~\ref{sca}(a) and (b)  we plot our results following the scaling relation. 
The points  between $\sim 18~$K and $\sim 30~$K collapse roughly onto a common curve.
A clear deviation for this scaling is observed below 18~K. We note that the 
predicted crossover at $x_2 \sim 1$ is well reproduced by the data.
 On the other hand, Hirschfeld et al. \cite{kub1,kub2,hir} showed that when one 
treats transport quantities, an exact one-parameter scaling is not necessarily obtained. In 
fact the account for an impuriy bandwith destroys the scaling properties and a plot of  
$\kappa_{xy}/T^2$ versus the scaling variable $x$ would only yield an approximate scaling 
in the best case. Either when the impurity band width becomes comparable to the magnetic energy or 
the temperature or the impurity relaxation rate becomes comparable to the vortex relaxation or 
inelastic collision rate, scaling should completely break down. 
This was also pointed out by Won and Maki \cite{won} although in this work the 
scaling law is recovered in the superclean limit $(\Gamma/\Delta \ll B/B_{c2} \ll  1, \Gamma$ is 
scattering rate due to impurities in $\hbar$ units). 
In the clean limit $(B/B_{c2} \ll   \Gamma/\Delta \ll 1)$ it was shown that the 
scaling properties break down again. If the used separation procedure to obtain the
electronic longitudinal thermal conductivity is really working properly, it should show the 
same scaling properties as the Hall conductivity. According to Ref.~\cite{sim1} the electronic thermal 
conductivity $\kappa^{el}_{xx}$ should show a scaling of the form $\kappa^{el}_{xx}/T \propto 
f_{xx}(x)$ with $x \propto \sqrt{B}/T$. The results in Fig.~\ref{keescal} indicate that $\kappa^{el}_{xx}$ 
shows the same scaling properties 
as $\kappa_{xy}$ in the same temperature range.

In conclusion, we think that our results for the electronic part of the longitudinal and the 
Hall component of  the thermal conductivity on a Y123 crystal do  show a convincing scaling 
in a restricted temperature range only. While the experimental points between 30 K and 18 K 
collapse roughly onto a common curve, the ones below and above this range spread out. 
This result partially contradicts the experimental results shown in Ref. \cite{sim2} 
since in that work no data were taken 
below 20 K. On the other hand the picture of an approximate scaling proposed first by 
K\"ubert and  Hirschfeld \cite{kub1} and later by Won and Maki \cite{won} appears 
to be consistent with our experimental data implying that an 
additional scattering/quasiparticle 
density or effective mass change
should be taken into account 
for the calculation of transport properties on these systems.

The observed upturn in the $c$-axis conductivity at $\sim 20~$K, see Fig.\ref{fig_tau}, 
 is not yet understood.\cite{hos,xiang} 
Because the c-axis conductivity decreases with temperature at $T < T_m$ it seems
reasonable to conclude that the accompanied anomaly in $\theta_H(T)$ is not related
to the crossover from holon
non-drag to a localization
regime \cite{anhall2}. 
Experimental results indicate that below $T_c$ the $c$-axis conductivity 
shows a clearly different behavior as the $ab$-plane conductivity,  an experimental fact
used as evidence for incoherent transport between CuO$_2$ planes \cite{hos}. 
A possible crossover to coherent transport at $T_m$ is, however, not yet clarified.\cite{xiang}
Ioffe and Millis \cite{ioffe} pointed out that the $c-$axis conductivity involves mainly 
the scattering mechanism of electrons within a CuO$_2$ plane. 
From this point of view appears reasonable to search for a common origin of the
anomaly at $T_m$ observed in the $c-$axis conductivity and the Hall signal.
However, this appears to be in contradiction to the different behavior 
of the electrical transport properties cited above. 
This apparently contradictory behavior may be related to 
the influence of the interlayer hopping integral on the $c-$axis transport
properties. According to Xiang and Hardy \cite{xiang}
the anisotropy of this hopping integral affects in such a way the $c-$axis conductivity that
it does not depend on the quasiparticle scattering rate and an universal $T^3$ dependence
below $T_c$ is obtained in agreement with the experimental results. 

Electrical Hall angle measurements in different HTS \cite{xu,abe,konst} show a clear 
 minimum in $\cot(\theta_H)$ above $T_c$ which has been identified as the
opening of the pseudogap. Therefore, we may speculate that 
this opening occurs at $T_m$ where $\cot(\theta_H)$ shows a clear minimum, see Fig.~\ref{fig_tau}. 
We note that the opening of a pseudogap {\em below} $T_c$  in Y123 HTS appears to be supported
by different experimental results.\cite{talo}
Supporting such a scenario, the opening of
a charge density wave gap at $T < 35~$K has recently been reported for Y123 crystal
with $T_c \simeq 90$~K \cite{kra}.
Does the opening of a pseudogap influence the scattering rate $\tau_H^{-1}$ or the effective mass
$m_H$ ? Since in a simple picture one expects the decrease of the scattering rate 
when the pseudogap opens, it has been suggested that the pseudogap affects
$\cot(\theta_H)$ through a change in the effective mass which may be related to
a modification in the Fermi surface topology.\cite{abe} 
The correlation between Hall angle and c-axis conductivity shown in
the present work would suggest that
the pseudogap influences the c-axis conductivity as well. If the scattering rate independence of 
the $c-$axis conductivity \cite{xiang}
would remain  valid at and below $T_m$, the correlation implies that the a similar mechanism that 
changes the effective mass at $T_m$ would be responsible for the upturn in the $c$-axis conductivity. 
We note, however,
that experimental results in Bi$_2$Sr$_2$CaCu$_2$O$_8$ HTS indicate that the opening of
the pseudogap reduces the $c-$axis conductivity in the normal state.\cite{laty} 
Future experiments should clarify if this behavior is also observed in Y123 HTS and below
$T_c$.

\section{Conclusion}

In summary,  we have measured the longitudinal and transverse thermal conductivity of 
a twinned and nearly optimally doped Y123 crystal as a function of temperature at different applied fields.
We have used two different approaches to derive the temperature dependence of the
Hall angle. Independently of the approach, we observe that 
$\cot(\theta_H)$ does not follow a $T^2$ dependence but a $T^4$ dependence at $T/T_c > \sim 0.3$, 
reaching a minimum 
at $T \sim 20~$K. The $T^4$ dependence agrees with the dependence of the scattering rate obtained
from longitudinal conductivity measurements.\cite{yu1}
The anomaly of $\cot(\theta_H)$ observed at 20~K
resembles that of the $c$-axis conductivity of a optimally doped crystal with 
similar $T_c$. Based on the proved sensitivity of the Hall angle to the 
pseudogap \cite{xu,abe,konst} we may speculate that this behavior  reflects the opening  of the pseudogap
at temperatures much below $T_c$. 
 We showed also that the electronic longitudinal and the thermal Hall conductivity show
a scaling only in a restricted temperature range. This scaling breaks down at temperatures
$T < 18~$K, where the Hall angle changes its behavior, and at $T > 30~$K.

\section*{ACKNOWLEDGMENTS}
We are specially grateful to B. Zeini, W. Hardy, A. G. Gaicochea, J. Kirtley,  K. Maki, A. Freimuth,  
Yu. Pogorelov, M. Ramos and S. Vieira for
informative discussions.
This work is supported by the Deutsche 
Forschungsgemeinschaft under DFG Es 86/4-3 and was partially supported by the
DAAD.

\end{document}